\documentclass[prc,showpacs,preprintnumbers,showkeys,
superscriptaddress]{revtex4}

\usepackage{enumerate} 
\usepackage{amsmath}
\usepackage{amsthm}
\usepackage{amsfonts}
\usepackage{amssymb}
\usepackage{amsxtra}
\usepackage{amssymb}
\usepackage{delarray}
\usepackage{graphicx}
\usepackage{dcolumn}
\usepackage{epsfig}
\usepackage{float}
\usepackage{hhline}
\restylefloat{figure}

\begin{document}

\title{Properties of asymmetric nuclear matter in different approaches}
\author{P. G\"{o}gelein}  
\affiliation{Institut f\"{u}r Theoretische Physik,
Universit\"{a}t T\"{u}bingen, D-72076 T\"{u}bingen, Germany}
\author{E.N.E. van Dalen}
\affiliation{Institut f\"{u}r Theoretische Physik,
Universit\"{a}t T\"{u}bingen, D-72076 T\"{u}bingen, Germany}
\author{Kh. Gad}
\affiliation{Physics Department, Faculty of Science, Sohag University,
Sohag, Egypt}
\author{Kh. S. A. Hassaneen}
\affiliation{Physics Department, Faculty of Science, Sohag University,
Sohag, Egypt}
\author{H. M\"{u}ther}
\affiliation{Institut f\"{u}r Theoretische Physik,
Universit\"{a}t T\"{u}bingen, D-72076 T\"{u}bingen, Germany}




\begin{abstract}
Properties of asymmetric nuclear matter are derived from various many-body
approaches. This includes phenomenological ones like the Skyrme
Hartree-Fock and relativistic mean field approaches, which are adjusted to fit
properties of nuclei, as well as more microscopic attempts like the
Brueckner-Hartree-Fock approximation, a self-consistent Greens function method
and the so-called $V_{lowk}$ approach, which are based on realistic nucleon-nucleon
interactions which reproduce the nucleon-nucleon phase shifts. These microscopic
approaches are supplemented by a density-dependent contact interaction to
achieve the empirical saturation property of symmetric nuclear matter.  
The predictions of all these approaches are discussed for nuclear matter
at high densities in $\beta$-equilibrium. Special attention is paid to behavior
of the isovector component of the effective mass in neutron-rich matter.
\end{abstract}

\keywords{Nuclear equation of state, isospin dependence, symmetry energy,
effective mass.}

\pacs{21.60.Jz, 21.65.-f, 26.60.-c, 97.60.Jd}
\maketitle

\section{Introduction}
The equation of state for nuclear matter under exotic conditions is one of the
main topics in modern nuclear physics. This interest in the properties of nuclear
matter at high densities and large proton-neutron asymmetries is partly 
motivated by the fact that this information is required in theoretical models of
compact objects in astrophysics like neutron stars or the simulation of
supernovae. However, the study of nuclear systems with large isospin asymmetries
are also the subject the forthcoming radioactive-ion-beam facilities such as FAIR
at GSI or SPIRAL2 at GANIL. 

Different ways have been developed to obtain predictions for the properties of
nuclear systems under exotic conditions. One way is to start from
phenomenological models which successfully describe the properties of stable
nuclei. A very popular approach along this line is the use of an effective
density dependent Skyrme-type interaction\cite{sk1,sk2,bv81,RS80,Rein07}. Modern
Skyrme parameterizations have been developed, which were constrained in their
fitting procedures to obtain results for neutron-rich nuclear matter compatible
to those of microscopic calculations. Here we mention the Skyrme-Lyon (SLy)
forces\cite{Chabanat}, which have been used in studies of the neutron-star
crust\cite{haensel,Goe07}.

Also the relativistic mean-field approximation has very successfully been used
to describe the properties of stable nuclei\cite{ring05}. Attempts have been
made to derive a set of density-dependent meson-nucleon coupling constants of
density dependent relativistic mean field (DDRMF) calculations from microscopic
Dirac Brueckner Hartree Fock calculations with a readjustment in order to
reproduce the bulk properties of stable nuclei and the saturation point of
symmetric nuclear matter\cite{Goe08}.

The parameters of such Skyrme Hartree Fock and DDRMF calculations have been
fitted to the data of stable nuclei and the predictive power of these simple
phenomenological approaches could be rather limited. This would mean that the
predictions for nuclear systems with exotic values for density and
proton-neutron asymmetries may not be very reliable. 

Therefore we will also consider some so-called microscopic approaches, which
start from models of the nucleon-nucleon (NN) interaction, which are adjusted to
describe the experimental phase shifts of NN scattering at energies below the
pion threshold\cite{mach1,arg1}. The traditional models of such realistic NN
interactions like e.g. the charge-dependent Bonn potential CDBONN\cite{mach1} or
the Argonne potential V18\cite{arg1} contain rather strong short range
components, which make it inevitable to employ non-perturbative approximation
schemes for the solution of the many-body problem for the nuclear hamiltonian
based on such interactions\cite{Muether00}. Such non-perturbative approximations
include the Brueckner hole-line expansion with the Brueckner Hartree Fock (BHF)
approximation, the self-consistent evaluation of Green's function using 
the T-matrix 
approximation\cite{dick1,dick2,boz2,boz3,fri03,frick05,Hassan04,Gad07} 
(SCGF) and also variational  approaches using correlated basis
functions \cite{akmal1,fabro2,fan1}.

If such microscopic calculations would reproduce the properties of nuclear
systems under normal conditions, one could argue that a scheme which reproduce
the data of two nucleons in the vacuum (NN phase shifts) as well as nuclear data
at normal densities should also provide reliable results for nuclear systems at
densities beyond the saturation density of nuclear matter. Unfortunately,
however, such microscopic calculations fail to reproduce the saturation point of
symmetric nuclear matter or the bulk properties of finite nuclei with good
precision\cite{Muether00}. For example in nuclear matter such calculations yield
results for the saturation point, which are located on the so-called Coester
band\cite{Coester}, i.e. they either yield too little binding energy or a
saturation density well above the empirical value of $\rho_0$ = 0.16 fm$^{-3}$.

In recent years large progress has been made developing tools for
essentially exact calculations for nuclear system with mass number up to about
$A=12$\cite{pieper}. These calculations demonstrate that using realistic models
for the NN interaction in a non-relativistic hamiltonian precise results for
nuclear systems are obtained if and only if these realistic two-body
interactions are supplemented by a three-body force. 

Therefore we will consider in the present study the results of BHF and SCGF
calculations, adding a simple density-dependent contact interaction which is
adjusted to describe the saturation point of symmetric nuclear matter.

During the last years a different calculation scheme has evolved, which is also
based on the NN scattering data but tries to decouple the low- and high-
momentum components in the nuclear hamiltonian using renormalization group
methods\cite{bogner,bogner2,Bogner03,herget}. The interaction resulting for the low-momentum regime of
nuclear structure, which we will refer to as $V_{lowk}$ is rather soft, which
implies that non-perturbative tools of many-body theory can be
applied\cite{bozevlow}. Also it is very attractive, that $V_{lowk}$ turns out to
be essentially model-independent, if the cut-off for the low-momentum regime is
chosen appropriately.

However, evaluating the properties of nuclear matter by using $V_{lowk}$ in a
Hartree-Fock or BHF calculation, one does not obtain a saturation
point\cite{bogner2,Kuckei03}, the binding energy per nucleon increases with increasing
density and nuclear systems tend to collapse to high densities. This will be
compensated by adding a density-dependent contact interaction, which is adjusted
in the same fashion as for the BHF and SCGF cases discussed above.

Therefore we will apply two different schemes: two phenomenological methods
(Skyrme Hartree Fock and relativistic mean field DDRMF) employing
parameterizations, which tend to reproduce some features of microscopic
calculations and three microscopic approaches (BHF, SCGF and Hartree-Fock with
$V_{lowk}$), which are supplemented by a simple contact interaction to reproduce
the empirical saturation properties of symmetric nuclear matter. For all these
methods we study bulk features of nuclear matter at large densities and proton
neutron asymmetries. A comparison of the resulting values for nuclear
compressibility, symmetry energy, proton fraction in $\beta$ equilibrium and
effective masses, which characterize the density of states around the Fermi
energy, should provide information about the uncertainties in the extrapolation
of nuclear properties in exotic regions of density and asymmetry. Also we expect
some hints about the reliability and problems of the individual approaches.

After this introduction the section two shall briefly outline the different
approximation schemes which we are going to consider. The results are discussed
in section three and  the main conclusions are summarized in the final section.

\section{Different many-body approaches}

\subsection{Skyrme Hartree-Fock}

A very popular many-body approach in nuclear physics is the Skyrme-Hartree-Fock
approach which can be found e.g. in \cite{sk1,sk2,bv81,RS80}.
The Skyrme interaction leads to an energy functional 
\begin{equation}
   E=\int \mathcal{H}(\boldsymbol{r}) \, d\boldsymbol{r},
\end{equation}
where $\mathcal{H}$ is the Hamiltonian density in the Hartree-Fock approximation.
In case of infinite asymmetric nuclear matter the Hamiltonian density
writes \cite{sk2,Chabanat}
\begin{equation}
 \mathcal{H} = 	\mathcal{H}_K + \mathcal{H}_{\text{eff}} + \mathcal{H}_0
 		+ \mathcal{H}_3   \label{eq:sk1}
\end{equation}
where $ \mathcal{H}_K $ is the kinetic energy term,
$\mathcal{H}_{\text{eff}}$ an effective mass term, 
$\mathcal{H}_0 $ a zero range term and $ \mathcal{H}_3 $ a density dependent term.
These terms are given by
\begin{eqnarray}
\mathcal{H}_K 	
	& = & \frac{ \hbar^2}{2m} \tau,  \notag \\
\mathcal{H}_{\text{eff}}	
	& = & \textstyle{\frac{1}{8}} 
		\big[ t_1 (2 + x_1 ) + t_2 (2 + x_2 ) \big] \tau \rho \notag\\
	&   & + \textstyle{\frac{1}{8}} \big[ t_2 ( 2x_2 +1 ) - t_1 ( 2 x_1 + 1 ) \big] 
		\big[ \tau_p \rho_p + \tau_n \rho_n \big], \notag \\
\mathcal{H}_0 	
	& = & \textstyle{\frac{1}{4}} t_0 
	      \big[ (2+x_0) \rho^2 - ( 2x_0 + 1 ) (\rho_p^2 + \rho_n^2 ) \big],  \notag \\
\mathcal{H}_3 	
	& = & \textstyle{\frac{1}{24}} t_3 \rho^\alpha 
	      \big[ (2 + x_3 ) \rho^2 - ( 2x_3 + 1 ) ( \rho_p^2 + \rho_n^2 ) \big],
\end{eqnarray}
where the coefficients $t_i$, $x_i$, $W_0$, and $\alpha$ are the parameters of a 
generalized Skyrme force. In the present study we apply the 
commonly used parameterization SLy4.

The densities $\rho$ and $\tau$ are defined in terms of the
corresponding densities for protons and neutrons 
$\rho = \rho_p + \rho_n$ and $\tau = \tau_p + \tau_n$. 
If we identify the isospin label $(i = p, n)$, 
the matter densities for protons and neutrons are given by
\begin{equation}
\rho_i 	= 2 \int_0^{k_{Fi}} \frac{d^3k}{(2\pi)^3} 
	=  \frac{1}{3 \pi^2} k_{Fi}^3,
\end{equation}
where $k_{Fi}$ denotes the corresponding Fermi momentum
and the spin degeneracy gives a factor of $2$.
We obtain the kinetic energy density  by
\begin{eqnarray}
\tau_i 	= 2 \int_0^{k_{Fi}} \frac{d^3k}{(2\pi)^3} k^2 
	= \frac{1}{5 \pi^2}(3\pi^2 \rho_i)^{5/3}.
\end{eqnarray}
The Hamiltonian density leads to the energy per nucleon
$E/A = \mathcal{H}/\rho$ 
and the single particle energy may be written as a function of momentum $k$
\begin{equation}
  \varepsilon_i(k) = \frac{k^2}{2m_i^\ast} + V_i
\end{equation}
with an effective mass $m_i^\ast$ and a density dependent Potential $V_i$.

\subsection{Brueckner-Hartree-Fock}

Starting from realistic NN interactions, we have to use more advanced 
many-body approximations like the Brueckner-Hartree-Fock (BHF) approach, 
which have the capability to account for the effects of correlations, which are
due to the strong tensor and short-range components of such realistic NN
interactions. In the BHF approximation this is achieved by evaluating the
so-called $G$-matrix, which corresponds to the in-medium scattering matrix.
The self-energy of a nucleon with isospin $i$, momentum $\boldsymbol{k}$
and energy $\omega$ in asymmetric nuclear matter is defined in the
BHF approximation by \cite{Muether00, Hassan04}
\begin{equation}\label{eq:SigmaBHF}
  \Sigma_i^{BHF} = \sum_{j} 
    \int d^3q \,
    \langle \boldsymbol{kq}  | G(\Omega) | \boldsymbol{kq} \rangle_{i j}
    n^0_{j}(\boldsymbol{q}).
\end{equation}
In this equation $n^0_{j}(\boldsymbol{q})$ refers to the occupation probability 
of a free Fermi gas of protons $(j=p)$ and neutrons $(j=n)$ like in the
mean-field or Hartree-Fock approach. This means that for asymmetric matter 
with a total density $\rho = \rho_{p} + \rho_{n}$
this probability is defined by
\begin{equation}
 n^0_{j}(\boldsymbol{q}) = 
	\begin{cases}  1 & \text{for} |\boldsymbol{q}|\leq k_{Fj}, \\
                       0 & \text{for} |\boldsymbol{q}| > k_{Fj},  \end{cases}
\end{equation}
with Fermi momenta for protons $(k_{Fp})$ and neutrons $(k_{Fn})$.

The antisymmetrized G matrix elements in eq.~(\ref{eq:SigmaBHF}) are obtained
from a given NN interaction by solving the Bethe-Goldstone equation
\begin{equation}\label{eq:BeGo}
\begin{split}
\langle \boldsymbol{kq}  | G(\Omega) | \boldsymbol{kq} \rangle_{ij}
  = &\langle \boldsymbol{kq}  | V | \boldsymbol{kq} \rangle_{ij}
     + \int d^3p_1 \, d^3p_2 \,
      \langle \boldsymbol{kq}  | V | \boldsymbol{p_1p_2} \rangle_{ij}   \\
    & \times \frac{Q(p_1i,p_2 j)}
              {\Omega - (\varepsilon_{p_1,i}+\varepsilon_{p_2,j}) + i\eta}  \\
    & \times \langle \boldsymbol{p_1p_2}  | G(\Omega) | \boldsymbol{kq}
    \rangle_{ij}.
\end{split}
\end{equation}
The single-particle energies $\varepsilon_{pi}$ of the intermediate states
should be the corresponding BHF single-particle energies which are defined
in terms of the real part of the BHF self-energy of eq. (\ref{eq:SigmaBHF}) by
\begin{equation}
  \varepsilon_{ki}=\frac{k^2}{2m}
	+\text{Re}\left[
	\Sigma_i^{BHF}(\boldsymbol{k},\omega=\varepsilon_{ki})\right],
\end{equation}
with a starting energy parameter $\Omega=\omega + \varepsilon_{qj}$ in the
Bethe-Goldstone equation (\ref{eq:BeGo}).

The Pauli operator $Q(p_1i,p_2j)$ restricts the intermediate 
states to particle states with momenta $p_1$, $p_2$,
which are above the corresponding Fermi momentum.
However, the single-particle spectrum is often parameterized in the
form of an effective mass
\begin{equation}
  \varepsilon_{ki} \approx \frac{k^2}{2m^\ast_i} + U_i,
\end{equation}
so that a so-called angle-averaged propagator can be defined,
which reduces the Bethe-Goldstone equation to an integral equation in one dimension.
The exact Pauli operator has been treated in \cite{Schill99}.

\subsection{Relativistic models}

An interesting extension of the BHF approach is the 
Dirac-Brueckner-Hartree-Fock (DBHF) approach which accounts for relativistic
effects as well as correlation effects as described in terms of  the G matrix.
In the DBHF approximation one evaluates the self-energy $\Sigma_i(k)$ in a way
very similar to the BHF approximation keeping track of the Lorentz structure. 
Attempts have been made to fit a density dependent 
Relativistic Mean Field model (DDRMF) to the results of such
DBHF calculations \cite{Fritz94,Goe08}.
Here we employ such a DDRMF model which has been obtained by fitting 
density dependent coupling constants
for the meson-nucleon vertices to reproduce the self-energy of the 
DBHF calculations of van Dalen et al. \cite{vD07}, which was based on the
Bonn A potential.

Both approaches, the DDRMF as well as the DBHF, start from a Lagrangian,
which includes baryons, mesons and the interaction 
\begin{equation}
  \mathcal{L} = \mathcal{L}_B + \mathcal{L}_M + \mathcal{L}_{\text{int}}.
\end{equation}
However, they differ in the included mesons and the coupling operators
\cite{Serot86,Brock90,Goe08,vD07}.
In the DDRMF model the $\sigma$, $\delta$, $\omega$, and $\rho$ mesons are
included in scalar and vector couplings, respectively.
The variation of the Lagrangian leads to a Dirac equation for the nucleons 
\begin{equation}\label{eq:Dirac}
 [\boldsymbol{\gamma} \cdot \boldsymbol{k_i^\ast} + m_{Di}^\ast] \,  u(k,s,i)
	= \gamma_0 \, E_i^\ast \, u(k,s,i),
\end{equation}
where $u(k,s,i)$ denotes the plane wave Dirac spinor with momentum $k$, 
spin $s$ and isospin $i$
\begin{equation}
  u(k,s,i) 
  	= \sqrt{ \frac{E^\ast_i + m^\ast_{Di}}{2E^\ast_i} }
  	  \begin{pmatrix}
	   1 \\
	   \frac{\boldsymbol{\sigma} \cdot \boldsymbol{k_i^\ast}}{E^\ast_i +
	   m^\ast_{Di}} 
	  \end{pmatrix}
	  \chi_{1/2}(s) \chi_{1/2}(i).
\end{equation}
The starred quantities contain the different components of the nucleon
self-energy: the scalar, time-like vector and space-like contributions
\begin{eqnarray}
 m_{Di}^\ast 		&=& M + \Sigma_{S,i}(k,k_F)	    \nonumber     \\
 E_i^\ast		&=& E_i(k) - \Sigma_{0,i}(k,k_F)      \nonumber     \\
 \boldsymbol{k_i^\ast} 	&=& \boldsymbol{k_i} + \boldsymbol{\hat{k_i}}
			    \, \Sigma_{V,i}(k,k_F),\label{eq:defrels}
\end{eqnarray}
where $\boldsymbol{\hat{k_i}}$ is the unit vector along 
the momentum $\boldsymbol{k_i}$ of the nucleon.
The general form of the nucleon self-energy in infinite matter
is obtained by evaluation of the meson exchange in spin saturated 
nuclear matter
\begin{equation}
 \Sigma_i(k,k_F) = \Sigma_{S,i}(k,k_F) + \gamma_0 \, \Sigma_{0,i}(k,k_F)
  + \boldsymbol{\gamma} \cdot \boldsymbol{\hat{k_i}}
    \, \boldsymbol{\Sigma}_{V,i}(k,k_F).
\end{equation}

In the fitting process special attention has to be payed to the 
rearrangement contribution to the time-like vector self-energy \cite{Fuchs95},
the relativistic operator structure \cite{Schill01}, and the proper
renormalization due to the space-like vector contribution to the self-energy
\cite{Goe08}.

The single particle energy in the DDRMF model is obtained from
the Dirac equation (\ref{eq:Dirac})
\begin{equation}
 E_i(k) = \sqrt{\boldsymbol{k_i^\ast}^2+ {m_{Di}^\ast}^2} +
 \Sigma_{0,i},\label{eq:dirace}
\end{equation}
and the energy-momentum tensor leads to the energy density in 
asymmetric nuclear matter
\begin{equation}
\begin{split}
\mathcal{E} = \langle T^{00} \rangle
	= &\frac{1}{\pi^2} \sum_{i=p,n} \int_0^{k_{F,i}} k^2 \, dk \,
		\sqrt{ \boldsymbol{k_i^\ast}^2 + m_{Di}^{\ast^2} }	\notag	\\
	& + \frac{1}{2} \sum_{i=p,n} 
	    \left( \Sigma_{S,i} \, \rho_i^s + \Sigma_{0,i} \, \rho_i \right),
\end{split}
\end{equation}
with the scalar densities $\rho_i^s$ and the baryon densities $\rho_i$
\begin{eqnarray}\label{eq:Reldens}
  \rho^s_i &=& \frac{1}{\pi^2} \int_0^{k_{F,i}} k^2 \, dk \,
	\frac{m_{Di}^\ast}{E^\ast_i(k)},  \notag \\
  \rho_i &=& \frac{1}{\pi^2} \int_0^{k_{F,i}} k^2 \, dk.
\end{eqnarray}

\subsection{Self-consistent Green's function} 

One of the drawbacks of the BHF approximation is the fact that it does not
provide results for the equation of state, which are consistent from the point
of view of thermodynamics. As an example we mention that BHF results do not
fulfill e.g. the Hugenholtz van Hove theorem. This is due to the fact that the
BHF approximation does not consider the propagation of particle and hole states
on equal footing. An extension of the BHF approximation, which obeys this
symmetry is the self-consistent Green's function (SCGF) method using the
so-called $T$-matrix approximation. During the last years techniques have been
developed, which allow to evaluate the solution of the SCGF equations for
microscopic NN interactions\cite{dick2,boz2,boz3,fri03,frick05}. Those
calculation demonstrate that for the case of realistic NN interactions, the 
contribution of particle-particle ladders dominates the contribution of
corresponding hole-hole propagation terms. This justifies the use of the BHF
approximation and a procedure, which goes beyond BHF and accounts for hole-hole
terms in a perturbative way\cite{Grange87,Frick02}. This leads to a modification
of the self-energy in the BHF approximation by adding a hole-hole term of the
form
\begin{equation}\label{eq:Sigma2h1p}
\begin{split}
  \Delta \Sigma_i^{2h1p}(k,\omega)
   =& \sum_{j} \int_{k_{Fj}}^\infty d^3p \int_0^{k_{Fi}} d^3h_1
      \int_0^{k_{Fj}}d^3h_2			\\
    & \times \frac{\langle \boldsymbol{kp} | G(\Omega) | \boldsymbol{h_1h_2} 
    \rangle_{ij}^2}
           {\omega+\varepsilon_{pj}-\varepsilon_{h_1i}-\varepsilon_{h_2j}-i\eta}.
\end{split}
\end{equation}
The quasi-particle energy for the extended self-energy can be defined as
\begin{equation}
  \varepsilon_{ki}^{qp}=\frac{k^2}{2m}
    +\text{Re}[\Sigma_i^{BHF}(k,\omega=\varepsilon_{ki}^{qp})
               +\Delta \Sigma_\tau^{2h1p}(k,\omega=\varepsilon_{ki}^{qp})],
\end{equation}
Accordingly, the Fermi energy is obtained evaluating this definition at
the Fermi momentum $k=k_{Fi}$ for protons and neutrons, respectively,
\begin{equation}
   \varepsilon_{Fi} = \varepsilon_{k_Fi}^{qp}.
\end{equation}

The spectral functions for hole and particle strength,
$S^h_i(k,\omega)$ and $S^p_i(k,\omega)$,
are obtained from the real and imaginary part of the
self-energy $\Sigma = \Sigma^{BHF} + \Delta \Sigma^{2h1p}$
\begin{equation}
  S^{h(p)}_i(k,\omega) =  \pm \frac{1}{\pi}
  \frac{\text{Im}\,\Sigma_i(k,\omega)}
       {[\omega-k^2/2m - \text{Re}\, \Sigma_i(k,\omega)]^2
        +[\text{Im}\, \Sigma_i(k,\omega)]^2}\,,				\\
\end{equation}
where the plus and minus sign on the left-hand side of this equation refers to
the case of hole ($h$, $\omega<\varepsilon_{Fi}$) and particle states ($p$, 
$\omega>\varepsilon_{Fi}$), respectively.
The hole strength represents the probability that a nucleon with isospin $i$,
momentum $k$, and energy $\omega$ can be removed from the ground state 
of the nuclear system with the removal energy $\omega$, 
whereas the particle strength denotes the probability
that such a nucleon can be added to the ground state of the system with $A$
nucleons resulting in a state of the $A+1$ particle system which has an energy
of $\omega$ relative to the ground state of the $A$ particle system.
Hence the occupation probability is obtained by integrating
the hole part of the spectral function
\begin{equation}
  n_i(k) = \int_{-\infty}^{\varepsilon_{Fi}}
    d\omega \, S^h_i(k,\omega).
\end{equation}
Note that this yields values for the occupation probability, which ranges
between values of 0 and 1 for all momenta $k$, leading to a partial depletion of
the hole-states in the Fermi gas model ($k<k_F$) and partial occupations for
states with momenta $h>k_F$.
A similar integral yields the mean energy for the distribution of the hole 
and particle strength, respectively
\begin{eqnarray}
  \langle \varepsilon_{hi}(k) \rangle
  &=& \frac{\int_{-\infty}^{\varepsilon_{Fi}} d\omega \,\omega \,
            S^h_i(k,\omega)}{n_i(k)},	\\
\langle \varepsilon_{pi}(k) \rangle
  &=& \frac{\int_{\varepsilon_{Fi}}^{\infty} d\omega \,\omega \,
            S^p_i(k,\omega)}{1-n_i(k)}.
\end{eqnarray}
Our self-consistent Green's function calculation is defined by identifying
the single particle energy in the Bethe-Goldstone equation as well as in 
the $2h1p$ correction term in eq.~(\ref{eq:Sigma2h1p}) by
\begin{equation}
  \varepsilon_{k\tau} = 
    \begin{cases}
       \langle \varepsilon_{h\tau}(k) \rangle &  \text{for } k < k_{F\tau}, \\
       \langle \varepsilon_{p\tau}(k) \rangle &  \text{for } k > k_{F\tau}.
    \end{cases}
\end{equation}
This definition leads to a single particle Greens function, which is defined
for each momentum $k$ by just one pole at $\omega=\varepsilon_{k\tau}$.
Hence in the calculation of the self-energy the mean value of the spectral
distribution is considered.
However, the modified occupation of nucleons obtained by the spectral functions are not 
included in the calculation of the self-energies.
The total energy per nucleon is evaluated by
\begin{equation}
  \frac{E}{A} = \frac{\sum_i\int d^3k\int_{-\infty}^{\varepsilon_{Fi}}
                      d\omega\, S^h_i(k,\omega) (k^2/2m+\omega)/2}
                     {\sum_i \int d^3k \, n_i(k)}.
\end{equation}

\subsection{Renormalization of the NN interaction} 

It is very reasonable to assume that the long-range or low-momentum part of the
NN interaction is fairly well described in terms of meson-exchange, while
different (quark) degrees of freedom are getting important to describe the
short-rang or high-momentum components of the NN interaction. Therefore it is
quite attractive to disentangle these low-momentum and high-momentum components
from each other. This means that one tries to define a model space, which
accounts for the low-momentum degrees of freedom and renormalizes the
effective hamiltonian for this low-momentum regime to account for the effects of
the high-momentum parts, which are integrated out.

This concept of a model space and effective operators appropriately
renormalized for this model space has a long history in
approaches to the nuclear many-body
physics. As an example we mention the effort to evaluate effective operators to
be used in Hamiltonian diagonalization calculations of
finite nuclei. For a review on this topic
see e.g.~\cite{morten:04}. The concept of a model space for
the study of infinite nuclear matter was used e.g. by 
Kuo et al.\cite{kumod1,kumod2,kumod3}.

During the last years this concept has received a lot of attention and led to
the definition of the so-called $V_{lowk}$ interaction. One way to determine
this interaction is to follow the unitary-model-operator approach
(UMOA)\cite{Suz94}. We follow the usual notation and define a projection operator 
$P$, which projects onto the model space of two-nucleon wave functions with
relative momenta $k$ smaller than a chosen cut-off $\Lambda$. The operator 
projecting on the complement of this subspace is
identified by $Q$ and these operators satisfy the usual relations like $P+Q=1$,
$P^2=P$, $Q^2=Q$, and $PQ=0=QP$. It is the aim of the Unitary Model Operator
Approach (UMOA) to define a unitary transformation $U$ in such a way, that the
transformed Hamiltonian does not couple the $P$ and $Q$ space, i.e.
\begin{equation}
  QU^{-1}HUP = 0\,.\label{eq:umoa}
\end{equation}
The technique to determine this unitary transformation is
nicely been described by Fujii et al.\cite{fuji:04} (see also \cite{bozevlow}).
It leads to an effective hamiltonian $H_{eff} = h_0 + V_{eff}$, which contains
the term of the kinetic energy $h_0$ and an effective interaction $V_{eff}$
given by
\begin{equation}\label{eq:Vlowk1}
  V_{eff} = V_{lowk}= U^{-1}(h_0 + V) U - h_0\,.
\end{equation}
Diagonalising this effective hamiltonian in the low-momentum model-space, one 
obtains eigenvalues which are identical to the diagonalisation of the original
hamiltonian $h_0+V$ in the complete space. This means solving the Lipmann
Schwinger equation for NN scattering using this $V_{lowk}$ with an cut-off
$\Lambda$ yields the same phase shifts as obtained for the realistic interaction
$V$ without a cutoff. One finds that the resulting $V_{lowk}$ is essentially
independent of the underlying realistic interaction $V$, if is fitted to the
experimental phase shifts and if the cut-off $\Lambda$ is chosen around
$\Lambda$ = 2 fm$^{-1}$, which means that the model space includes scattering
up to the pion threshold. In that sense $V_{lowk}$ is unique and, as it
reproduces the NN scattering phase shifts it can also be regarded as a realistic
interaction like e.g. the CDBONN or Argonne V18 interactions.

Since, however, the high-momentum or short-range components have been integrated
out by means of the unitary transformation of eq.(\ref{eq:umoa}), the $V_{lowk}$
does not induce any short-range correlations into the nuclear wave function.
This leads to the nice feature that mean-field calculations using $V_{lowk}$ 
lead to reasonable results and corrections of many-body theories beyond mean
field are weak\cite{bozevlow}. On the other hand, however, it is this lack of
short-range correlation effects, which are modified in the medium, which
prevents the emergence of a saturation point in calculations of symmetric
nuclear matter\cite{bogner2,Kuckei03}. 

In order to to achieve saturation in nuclear matter one has to add three-body
interaction terms or a density-dependent two-nucleon interaction. This is not
very astonishing as it is known that a renormalization of a two-body operator
leads to many-body terms\cite{polls:83,polls:85}. Therefore it is quite natural
to supplement the effective interaction $V_{lowk}$ by a simple contact
interaction, which we have chosen following the notation of the Skyrme
interaction to be of the form
\begin{equation}
  \Delta\mathcal{H} = \frac{1}{2} t_0 \rho^2 + \frac{1}{12} t_3
  \rho^{2+\alpha}\,,\label{eq:contact}
\end{equation}
where $\rho$ is the matter density, $t_0$, $t_3$ and $\alpha$ are parameters.
For a fixed value of $\alpha$ (typically $\alpha$ =0.5) we have fitted $t_0$ and
$t_3$ in such a way that a Hartree-Fock calculation using $V_{lowk}$ plus the
contact term of eq.(\ref{eq:contact}) yields the empirical saturation point for
symmetric nuclear matter. 

The same parameterization of a contact term has been used to evaluate corrections
to the self-energy of BHF and SCGF in such a way that also these calculations
reproduce the saturation of symmetric nuclear matter. 

Note that this contact term is an isoscalar
term and does not influence the symmetry energy,
proton fractions in $\beta$-equilibrium, and effective masses.

\section{Results and Discussion}

All results of calculations, which refer to realistic NN interactions, have been
obtained using the CDBONN\cite{mach1} interaction. This includes all BHF and
SCGF calculations. Also the evaluation of $V_{lowk}$ has been based on the
proton-neutron part of CDBONN. Using a cut-off parameter $\Lambda$ = 2
fm$^{-1}$, these results do not significantly depend on the underlying
interaction. The Skyrme Hartree-Fock calculations have been done using the
parameterization SLy4 and for the relativistic mean-field calculation the
parameterization for DDRMF in \cite{Goe08} has been used.

\begin{figure}
\begin{center}
\mbox{
\includegraphics[width =10cm]{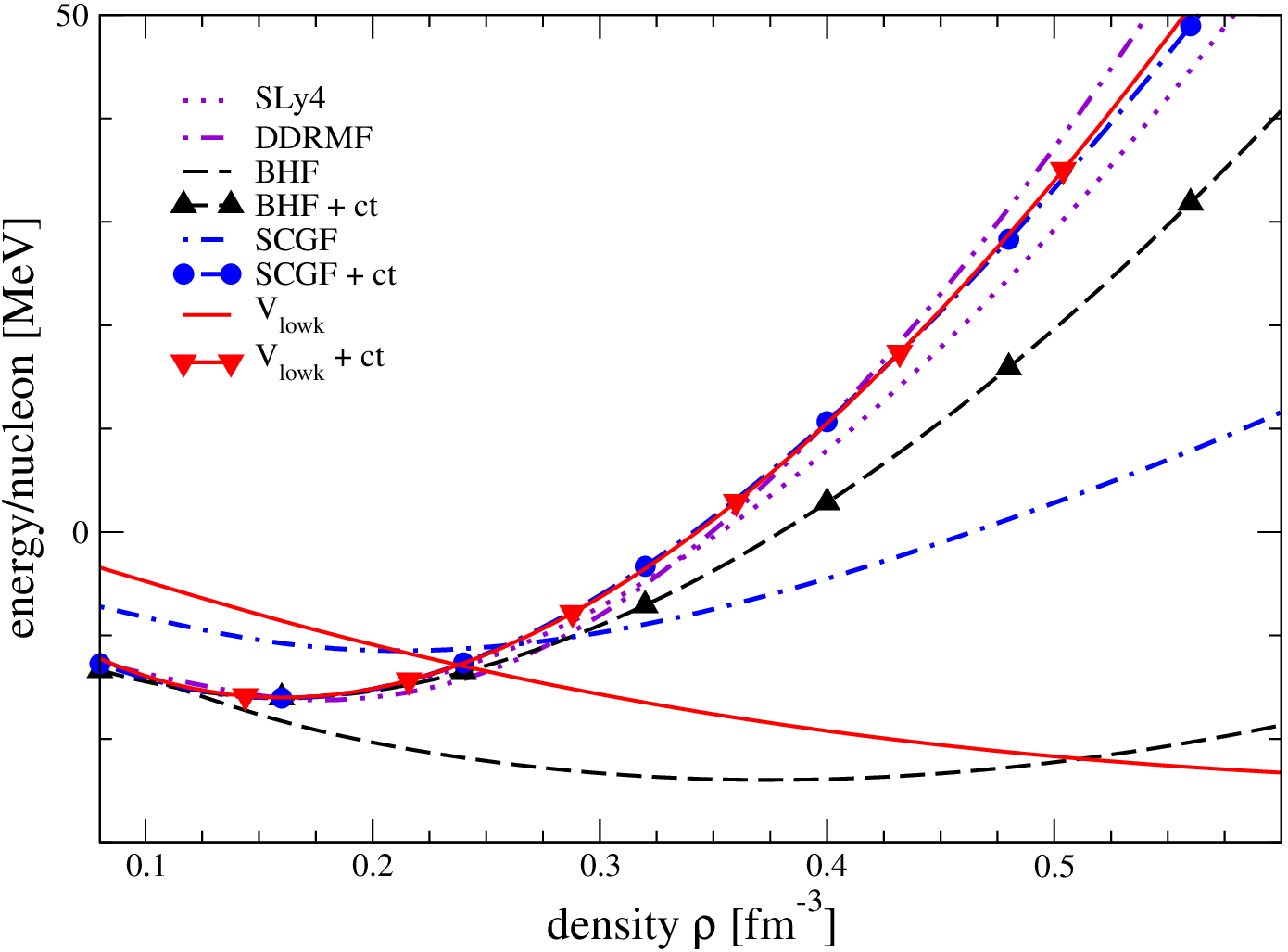}
}
\end{center}
\caption{\label{fig:CompSymM} (Color online) 
Comparison of binding energy per nucleon of symmetric nuclear matter 
as obtained from Skyrme SLy4, DDRMF, BHF, SCGF, and Vlowk.
Results of approaches based on realistic NN interactions
are also compared with an additional contact interaction of the form displayed
in eq.\protect{\ref{eq:contact}}}
\end{figure} 

First let us turn to the binding energy of symmetric nuclear matter, which are
displayed in Fig.~\ref{fig:CompSymM}. Compared to other realistic NN
interactions the CDBONN potential, which we have chosen here is a rather soft NN
interaction with a weak tensor force. This is indicated by the results for the
saturation point of symmetric nuclear matter as obtained in the BHF 
approximation (the minimum of the dashed black line in Fig.~\ref{fig:CompSymM}
and data in table~\ref{table:NM_Prop} ). The saturation density is larger than
twice the empirical value and the calculated energy is well below, which means
that the CDBONN result is located in the large binding energy high density part
of the Coester band\cite{Muether00}. 

\begin{table}
\begin{center}
\begin{tabular}{|ll|ccccccc|}
\hline
	&& SLy4 \ & DDRMF \ &BHF & BHF(+ct) & SCGF& SCGF(+ct)&  
	$V_{lowk}$ + ct   \\
\hline
&&&&&&&&\\
$\rho_0$ & [fm$^{-3}]$ & 0.160 & 0.178 & 0.374 & 0.161 & 0.212 & 0.160 & 0.160 \\
$E/A(\rho_0)$   & [MeV]	& -15.97 & -16.25 & -23.97 & -16.01 & -11.47 & -16.06 & -16.0  \\
$K$         & [MeV] & 230 & 337	& 286 & 214 & 203 & 270 & 258       \\
$a_S(\rho_0)$  & [MeV] & 32.0 & 32.1 & 51.4 & 31.9 & 34.0 & 28.3 & 21.7     \\ 
\hline
\end{tabular}
\end{center}
\caption{\label{table:NM_Prop}
Properties of symmetric nuclear matter are compared for Skyrme SLy4, DDRMF, BHF, SCGF,
and $V_{lowk}$. The results, which are listed in the columns labeled with +ct
are obtained employing the additional contact interaction of
eq.(\protect{\ref{eq:contact}}) with parameters as listed in
table~\protect{\ref{table:CT_Param}}.
The quantities listed include the saturation density $\rho_0$, the 
binding energy at saturation $E/A$, the compressibility modulus $K$ and the 
symmetry energy at saturation density $a_S(\rho_0)$.}
\end{table}

The hole-hole contributions to the nucleon self-energy, which are included in
the SCGF approach yield a repulsive contribution the energy per nucleon, which
increases with increasing nuclear density (dashed-dotted green line in
Fig.~\ref{fig:CompSymM}). This shifts the saturation point to a lower density
and binding energy per nucleon so that also the saturation point obtained with
SCGF is within the Coester band. 

The Hartree-Fock calculation using $V_{lowk}$ does not lead to a minimum in the
energy versus density plot (see solid red line in Fig.~\ref{fig:CompSymM}), as
we have already mentioned before\cite{bogner2,Kuckei03}. 

\begin{table}
\begin{center}
\begin{tabular}{|ll|ccc|}
\hline
	&& \ BHF \ \	& \ SCGF \ \	& \ $V_{lowk}$ \  \\
\hline
$t_0$\ &\ [MeV fm$^3$]	 	& -153	& -311 	& -438.1		\\
$t_3$\ &\ [MeV fm$^{3+3\alpha}$]	& 2720	& 3670	& 6248		\\
\hline
\end{tabular}
\end{center}
\caption{\label{table:CT_Param}
Parameters $t_0$ and $t_3$ defining the contact interaction of 
eq.(\protect{\ref{eq:contact}})as obtained for the fit to the saturation point
$\rho=0.16 fm^{-3}$ and $E/A=-16.0$ MeV at $\alpha=0.5$ for various 
realistic approaches.}
\end{table}

In order to reproduce the empirical saturation point of symmetric nuclear we
have added an isoscalar interaction term as defined in eq.(\ref{eq:contact})
choosing a value for $\alpha$ =0.5 and fitting the parameters $t_0$ and $t_3$.
The results for these fitting parameters are listed in
table~\ref{table:CT_Param} and the corresponding energy versus density
curves are displayed in Fig.~\ref{fig:CompSymM} using the same line shape and
color with and without adding the contact term but add an additional symbol to
the lines displaying the results with inclusion of the contact term. For all
three cases the fit yields an attractive two-body contact interaction and a
repulsive $t_3$ term.

The results for the calculated saturation points in table~\ref{table:NM_Prop}
are supplemented by the corresponding values for the nuclear compressibility 
modulus 
\begin{equation}
K= 9 \rho_0^2 \frac{\partial ^2 (E/A)}{\partial \rho^2}\bigg|_{\rho=\rho_0}
\end{equation}
This nuclear compressibility, which is calculated at the saturation density
$\rho_0$, 
together with the increase of energy at large density displayed in 
Fig.~\ref{fig:CompSymM} characterize the stiffness of the EoS of symmetric
nuclear matter. Comparing the different approaches we find that the relativistic
features included in the DDRMF approach lead to the stiffer EoS around the
saturation density as well as at higher densities. The SCGF and the $V_{lowk}$
calculations yield rather similar results after the contact terms are included,
which are a little bit softer than the DDRMF results and characterized by a
compression modulus of 270 MeV and 258 MeV for SCGF and $V_{lowk}$,
respectively. At higher densities the results are also very close to those
obtained for the Skyrme Hartree-Fock using SLy4. Note, however, that SLy4 yields
a rather low value for $K$ as compared to the SCGF and $V_{lowk}$ calculations.
The softest EoS for symmetric matter among those approaches which fit the
empirical saturation point is provided by the BHF approximation.

It is obvious that these results for the EoS for BHF, SCGF and $V_{lowk}$ are
rather sensible to the choice of the exponent $\alpha$ in the $t_3$ term of the
contact interaction. A larger value of $\alpha$ would lead to stiffer EoS. We
have picked the value $\alpha$ = 0.5 to obtain results for the EoS, which are
similar to those resulting from the empirical approaches. Note, that the choice
of $\alpha$ does not have any effect on the results referring to proton neutron
asymmetries, which is the main focus of this study. 

Table~\ref{table:CT_Param} also displays results for the symmetry energy
\begin{equation}
 a_S(\rho)= \frac{\partial (E/A)}{\partial I^2}\bigg|_\rho, 
 \qquad \quad I=\frac{N-Z}{A}=1-2Y_p,\label{eq:symm}
\end{equation}
evaluated for each approach at the corresponding saturation density $\rho_0$.
The two phenomenological approaches SLy4 and DDRMF yield results which are in
the range of the experimental value of $32\pm 1$ MeV. Also the BHF and SCGF
approach lead to results which are rather close to the empirical value, if the
contact term has been added. The BHF and SCGF calculations without the contact
term lead to non-realistic values for $a_S(\rho_0)$ since these values are
calculated at the corresponding saturation densities, which are larger than the
empirical saturation density. 

\begin{figure}
\begin{center}
\mbox{
\includegraphics[width =10cm]{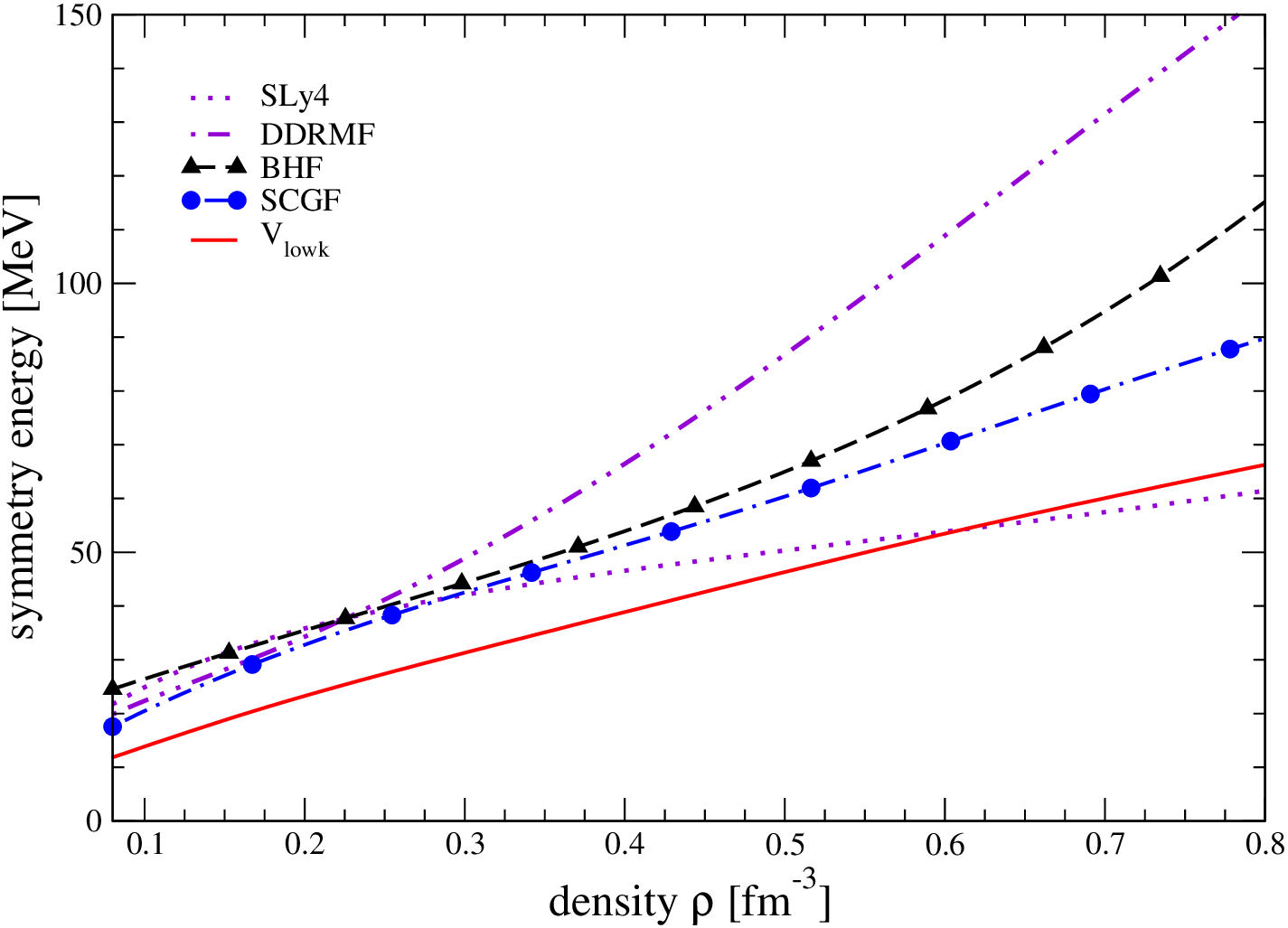}
}
\end{center}
\caption{\label{fig:aSym} (Color online) 
Comparison of the symmetry energy $a_S(\rho)$ as a function of density $\rho$
as obtained from Skyrme SLy4, DDRMF, BHF, SCGF, and $V_{lowk}$ approaches.}
\end{figure} 

\begin{figure}
\begin{center}
\mbox{
\includegraphics[width =10cm]{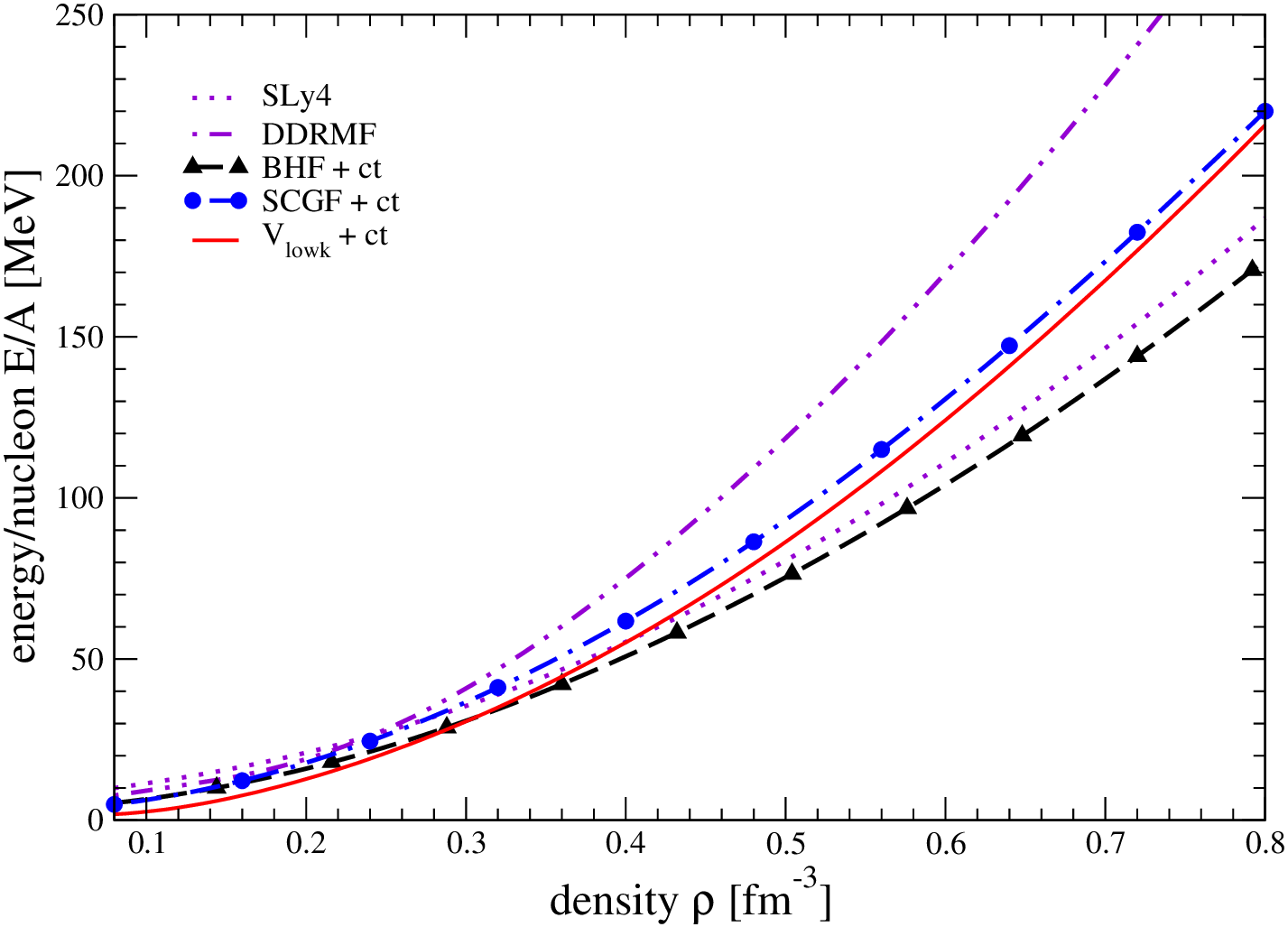}
}
\end{center}
\caption{\label{fig:pureN} (Color online) 
Energy per nucleon of pure neutron matter as a function of density
as obtained from Skyrme SLy4, DDRMF, BHF, SCGF, and $V_{lowk}$ approaches.}
\end{figure}

The symmetry energy calculated in the SCGF approach is slightly smaller than the
one obtained from the BHF approximation. This is valid for all densities under
consideration (see Fig.~\ref{fig:aSym}). This difference can easily be
explained: As we already mentioned above, the contribution of the hole-hole terms
is repulsive, which leads to larger energies for SCGF as compared to
BHF for all densities in symmetric nuclear matter (Fig.~\ref{fig:CompSymM}) as
well as in pure neutron matter (see Fig.~\ref{fig:pureN}). Since, however, the
contribution of ladder diagrams is larger in the proton-neutron interaction (due
to the strong tensor terms in the $^3S_1-^3D_1$ partial wave) than in the
neutron-neutron interaction, this repulsive effect is stronger in symmetric
nuclear matter than in neutron enriched matter. Therefore the symmetry energy
calculated in SCGF is slightly smaller if the hole-hole terms are included in
SCGF\cite{dieper03}. 

The symmetry energy at saturation density obtained from 
$V_{lowk}$ plus contact term is only about two third of the experimental value
(see table~\ref{table:NM_Prop}) and the value is significantly below the other
two microscopic approaches also at higher densities. This can be understood from
the following considerations: The Hartree-Fock calculations using $V_{lowk}$ do
not account for the attractive contributions due to the NN ladder terms involving
NN states with relative momenta below the cut-off $\Lambda$. This missing
attraction is compensated by the fit of the contact interaction to the empirical
saturation point of symmetric matter. While the contact interaction is chosen to
be identical for proton-neutron and neutron-neutron interaction, the ladder
terms are more attractive for the isospin $T=0$ partial waves (see above), 
i.e. the proton-neutron interaction. This leads to a significant underestimate 
for the symmetry energy at all densities. Note, however, that this failure of
$V_{lowk}$ should disappear if $V_{lowk}$ would be employed in an appropriate  
many-body calculation beyond the mean field approximation.

The symmetry energy rises as a function of density for all approaches
considered. Note, however, that the two phenomenological approaches Skyrme
Hartree-Fock using SLy4 and DDRMF provide rather different predictions at high
densities although the symmetry energy at normal density is identical. The
relativistic approach predicts symmetry energies for high densities, which are
well above all those derived from the microscopic calculations, while the Skyrme
interaction yields a symmetry energy which is even below the $V_{lowk}$ estimate
at densities above four times saturation density. 

\begin{figure}
\begin{center}
\mbox{
\includegraphics[width =10cm]{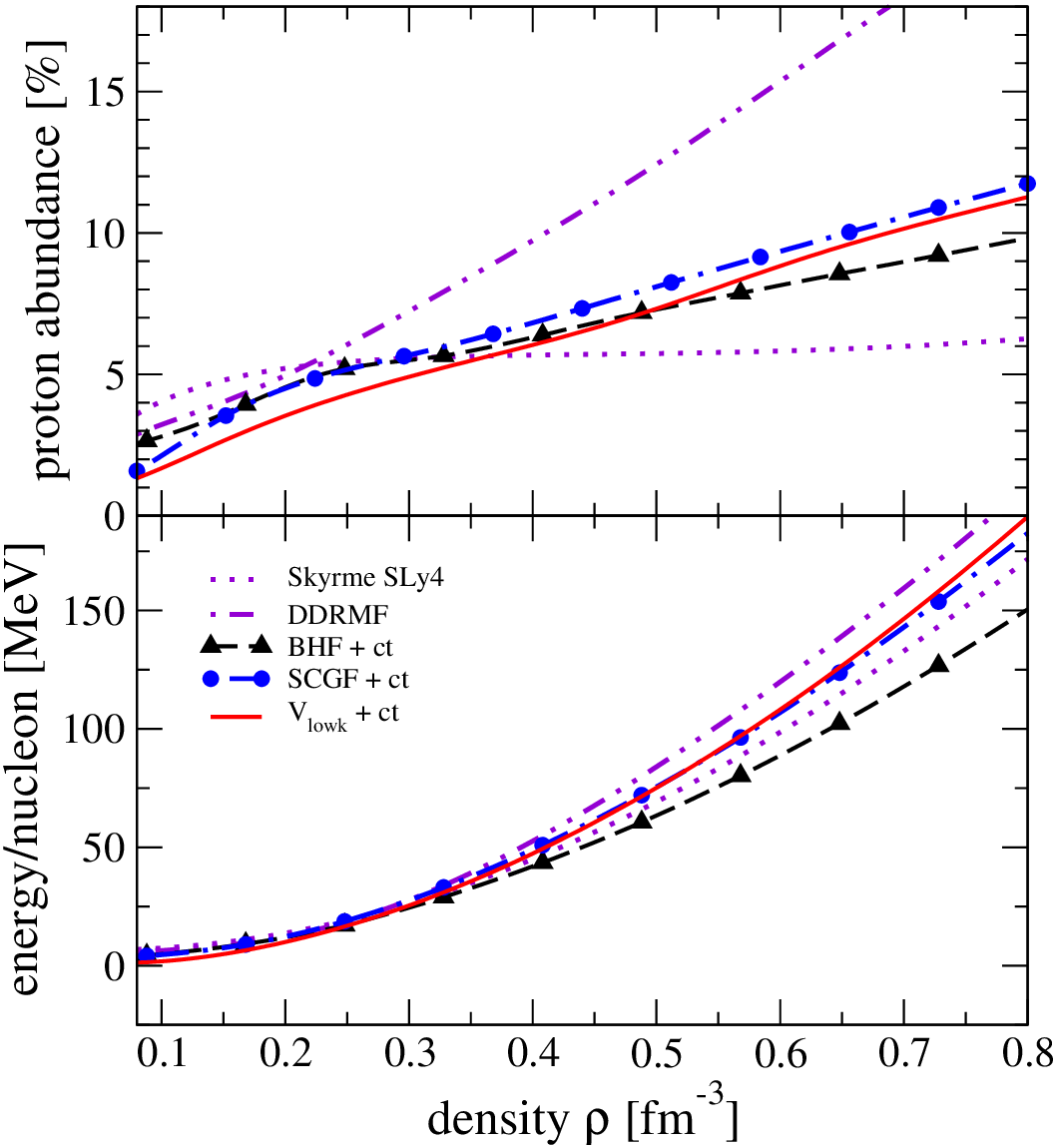}
}
\end{center}
\caption{\label{fig:betaEQ} (Color online) 
Results for a system of infinite matter consisting of protons, neutrons and
electrons in $\beta$-equilibrium.
The upper panel show the proton abundances and the lower panel displays the
energy per nucleon as a function of density using the various approximation
schemes discussed in the text.}
\end{figure}

Rather similar features also observed, when we inspect the properties
of nuclear matter in $\beta$-equilibrium, neutralizing the charge of the protons
by electrons, displayed in Fig.~\ref{fig:betaEQ}. The  upper panel of this
figure displays the proton abundance $Y_p= Z/A$, which are to some extent
related to the symmetry energy: large symmetry energy should correspond to large
proton abundances. So the largest proton abundances
are predicted within the DDRMF approach. Already at a density around 0.4 fm$^{-3}$
$Y_p$ exceeds the about 10\%, which implies that the direct URCA process could
be enabled, which should be reflected in a fast cooling of a neutron star.

The other extreme case is the prediction derived from SLy4. In this approach the
proton abundance does not exceed a value of 6\%.  

The $V_{lowk}$ and SCGF approaches lead to similar proton abundances at
large densities. This demonstrates that the evaluation of the proton abundance
in $\beta$-equilibrium cannot directly be deduced from the symmetry energy,
since the former observable is derived from proton and neutron energies at large
asymmetries ($Z<<N$), whereas the symmetry energy is calculated from the 
second derivative at $N=Z$ (see eq.(\ref{eq:symm}). 
The BHF approach shows slightly lower values for $Y_p$ at
high density, but the results are still in the same range as SCGF and $V_{lowk}$.

At low densities the Skyrme HF approach yields large proton fractions as
compared to the results of the other calculations.
Large proton fractions at low densities tend to enhance density inhomogeneities
and thus favor the existence of a large variety of pasta structures. Therefore 
the Skyrme HF (Sly4) and the DDRMF approach, which
have been explored in detail in \cite{Goe07,Goe08}, should favor the formation of 
pasta structures as compared to the microscopic approaches.

Comparing the energies of matter in $\beta$-equilibrium derived from the various
approaches as a function of density (Fig.~\ref{fig:betaEQ},lower panel)
we find the same trends as in the case of
pure neutron matter displayed in Fig.~\ref{fig:pureN}. The absolute values are
lower in the case $\beta$-equilibrium (about 75\% of the energies for neutron
matter). Furthermore the relative differences between
the various approaches are smaller. While in the case of neutron matter the
energy differences between the various predictions are as large as 50\% of the
mean value, the corresponding number for matter in $\beta$-equilibrium is only
around 25\%. The approximation schemes leading to large
energies for neutron matter also show large symmetry energies, which result into
relatively large proton abundances and smaller energies for matter in 
$\beta$-equilibrium. 

The equation of state of nuclear matter in $\beta$-equilibrium is the main input
to predict mass and radii of neutron stars. A stiffer equation of state supports
a larger maximum mass and a lower central density.
In addition a thicker crust is found for the stiffer equation of 
state\cite{Engvik96}.

\begin{figure}
\begin{center}
\mbox{
\includegraphics[width =10cm]{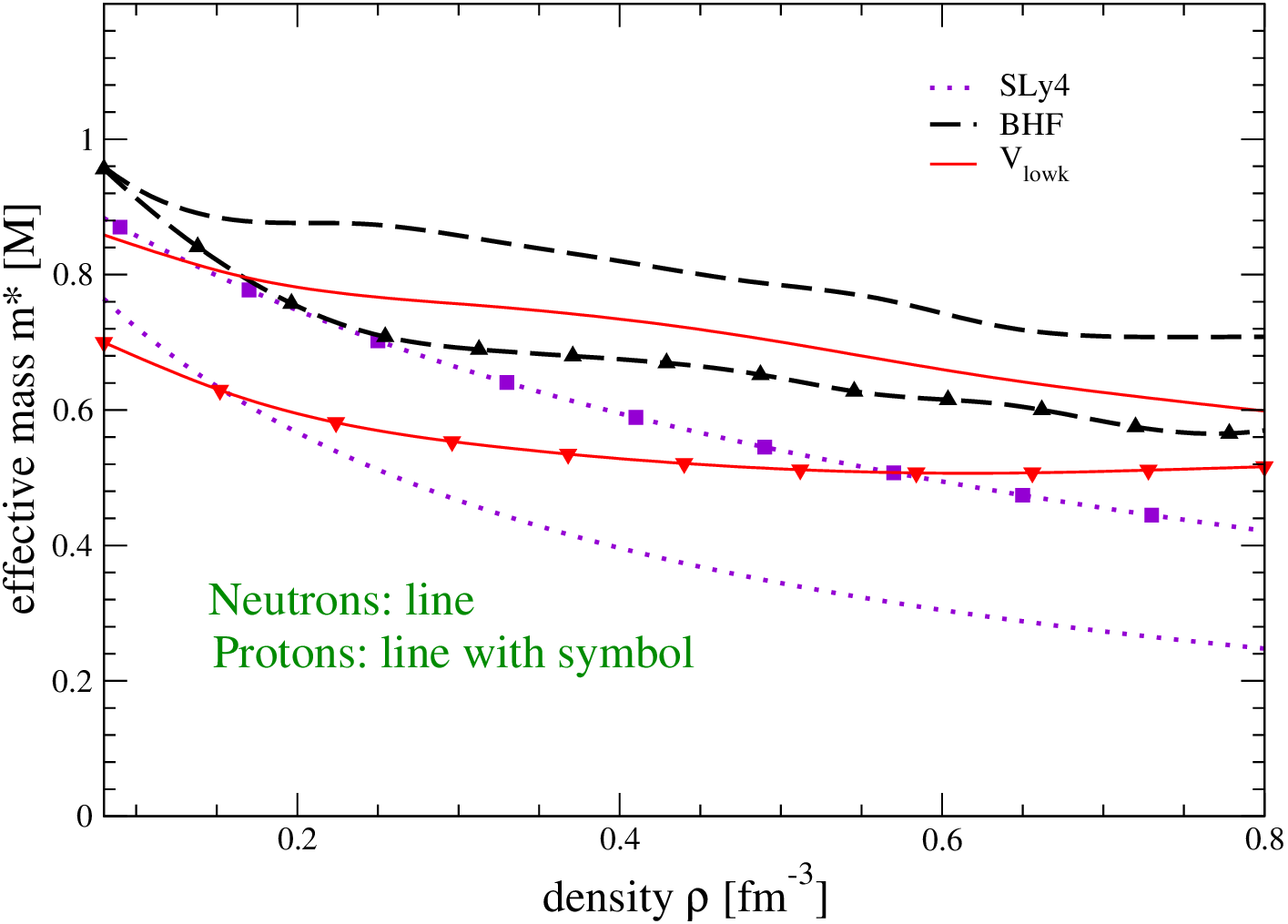}
}
\end{center}
\caption{\label{fig:effMass} (Color online) Effective masses for protons (lines
with symbols) and neutrons (lines without symbols) as obtained for nuclear
matter in $\beta$-equilibrium using 
Skyrme HF (SLy4), BHF, and $V_{lowk}$ approaches.}
\end{figure}

Another important information for the evaluation of dynamical features of matter
in neutron stars is the density of states,  which can be characterized by an
effective mass. The term effective mass is used in various connections in
many-body physics. This includes the effective Dirac mass of the relativistic
mean field approach $m_D^\ast$ (see eq.(\ref{eq:defrels})), as well as effective
masses, which express the non-locality of the self-energy in space and time,
which corresponds to a momentum and energy dependence. The density of states,
however, is related to the single-particle spectrum close to the Fermi energy, 
which in the case of nuclear matter can be parameterized in terms of an 
effective mass by the expression
\begin{equation}
  \varepsilon(k) = \frac{k^2}{2m^\ast} + U\,.\label{eq:defmas}
\end{equation}
Such effective masses for protons and neutrons determined for nuclear matter 
in $\beta$-equilibrium are
displayed  in Fig.~\ref{fig:effMass} as a function of density considering
non-relativistic approximation schemes.

It is a general feature of all approaches considered that the effective masses
for protons as well as neutrons decrease with increasing density. However, there
is a striking difference between the phenomenological Skyrme approximation and
the BHF and $V_{lowk}$ approach, which are based on realistic NN interactions:
The effective mass for protons is smaller than the corresponding one for
neutrons in neutron rich matter for the calculations using realistic
interactions, while it is opposite applying the Skyrme parameterization. In
fact,
if we define the effective masses for protons $m_p^\ast$ and neutrons
$m_n^\ast$ in terms of isoscalar $m_S^\ast$ and isovector masses $m_V^\ast$
by
\begin{eqnarray}
\frac{1}{m_n^\ast} & = & \frac{1}{m_S^\ast} + I \left(\frac{1}{m_S^\ast} -
\frac{1}{m_V^\ast}\right)\nonumber \\
\frac{1}{m_p^\ast} & = & \frac{1}{m_S^\ast} - I \left(\frac{1}{m_S^\ast} -
\frac{1}{m_V^\ast}\right)\nonumber \\
\mbox{with}\qquad I & = & \frac{N-Z}{A}\,,\label{eq:isomass} 
\end{eqnarray}
it turns out
most of the Skyrme parameterizations yield an effective isovector mass
$m_V^\ast$, which is even larger than the bare nucleon mass $M$\cite{Rein07},
which implies that it is larger than the effective isoscalar mass $m_S^\ast$.
This means 
that the effective mass for neutrons is smaller than the corresponding one for
the protons in neutron rich matter ($I>0$). These Skyrme parameterizations
leading to a large effective isovector mass are usually favored as they
correspond within the mean-field approach to an enhancement factor $\kappa$ of the
Thomas-Reiche-Kuhn sum-rule\cite{RS80,Bender03}. Attempts have been made to
distinguish between effective masses, which describe the energies around the
Fermi energy, and those characterizing the bulk spectrum by introducing a
term, which leads to a surface peaking of the effective mass term in finite
nuclei\cite{Farine01}.

\begin{figure}
\begin{center}
\mbox{
\includegraphics[width =8cm]{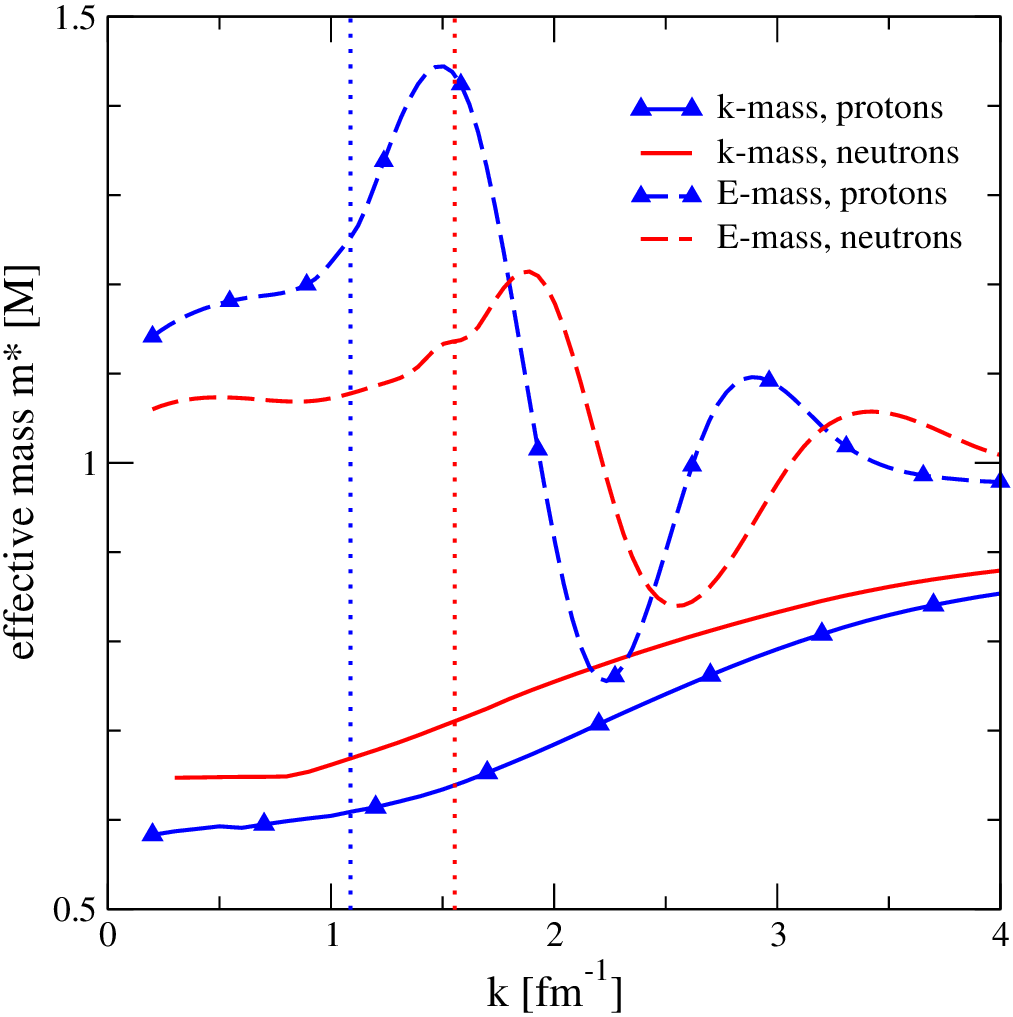}
}
\end{center}
\caption{\label{fig:MeffSCGF} (Color online)
Effective $k$-mass $m^\ast_k(k)$ (solid lines) and
effective $E$-mass $m^\ast_E(k)$ (dashed lines) for neutrons and protons (lines
with symbol) as obtained from the BHF calculations for asymmetric nuclear matter
at the density $\rho=0.17 \text{ fm}^{-3}$ and a proton abundance of 25 \%. The
Fermi momenta for protons and neutrons are indicated by vertical dotted lines}
\end{figure}

Non-relativistic descriptions of nuclear matter, which are based on realistic
interactions yield an effective isovector mass $m_V^\ast$ which is smaller
than the corresponding effective isoscalar mass, which leads to a larger
effective mass for neutrons than for protons in neutron-rich matter (see 
Fig.~\ref{fig:effMass}). In order to analyze this finding we inspect the
dependence of the nucleon self-energy in the BHF approximation
$\Sigma_i^{BHF}$, defined in eq.(\ref{eq:SigmaBHF}), as a function of energy
$\omega$ and momentum $k$ of the nucleon considered. Following  the discussion
of Mahaux and Sartor\cite{Mahaux91} one can define the effective $k$-mass
\begin{equation}
  \frac{m_k(k)}{M} = \left[ 1 + \frac{M}{k} 
    \frac{\partial \Sigma ( k, \omega) }{\partial k} \right]^{-1}
\end{equation}
and the effective $E$-mass
\begin{equation}
  \frac{m_E(\omega)}{M} = \left[ 1 -  
    \frac{\partial \Sigma ( k, \omega) }{\partial \omega} \right].
\end{equation}
The effective mass can then be calculated from the effective
$k$-mass and the effective $E$-mass by
\begin{equation}
  \frac{m^\ast(k)}{M} = \frac{m^\ast_k(k)}{M} \frac{m^\ast_E(\omega = 
  \varepsilon(k))}{M}.\label{eq:mstartot}
\end{equation}
Results for the effective $k$-mass and $E-mass$ as obtained from BHF
calculations for asymmetric nuclear matter at a density $\rho$ = 0.17 fm$^{-3}$
and a proton abundance $Y_\pi$ of 25 \% ($I=0.5$) are displayed in 
Fig.~\ref{fig:MeffSCGF}. We notice that the effective $k$-mass for the protons 
is significantly below the corresponding value for the neutrons at all momenta.
Since the $k$-masses tend to increase as a function of the nucleon momentum $k$,
the difference in the Fermi momenta for protons and neutrons enhance the
difference $m^\ast_{k,n} (k_{Fn}) - m^\ast_{k,p} (k_{Fp})$. 

The effective $k$-mass describes the non-locality of the BHF self-energy. This
non-locality and thereby also these features of the effective $k$-mass are
rather independent on the realistic interaction used. Furthermore it turns out
that the values for the $k$-mass are essentially identical if one derives them
from the nucleon BHF self-energy using the $G$-matrix or from the bare
interaction $V$ or from $V_{lowk}$\cite{Frick02}. This non-locality of the 
self-energy is dominated by Fock-exchange contribution originating from
$\pi$-exchange. In neutron-rich matter this contribution leads to a stronger
depletion for the proton mass than for the neutron mass\cite{Hassan04,Zuo05}. 

The effective $E$-mass, representing the non-locality of the self-energy in time, 
yield values larger than $M$ for momenta around $k_F$. Also in this case the
deviation of $m^\ast$ from $M$ is larger for protons than for neutrons. The
effective $E$-mass originates from the energy-dependence of the $G$-matrix and
is due to the admixture of 2-particle 1-hole configurations to the
single-particle states. One may also say that the effective $E$-mass is due to
correlations beyond mean field. It accounts to the coupling of vibrational
modes.
 
Anyway, the enhancement of the effective mass $m^\ast$, which is due to the
effective $E$-mass in eq.(\ref{eq:mstartot}) is not strong enough to compensate
the effects of the $k$-mass. Therefore the final effective mass is below the
bare mass $M$ and the effective mass for neutrons remains larger than the
corresponding one for protons.

The effects of the $E$-mass are weaker for $V_{lowk}$ than for the $G$-matrix.
This is obvious since $V_{lowk}$ only accounts for ladder diagrams included with
particle-particle states above the cut-off, whereas the $G$-matrix includes all
particle-particle states above the Fermi-momenta. This explains the lower
effective masses obtained for $V_{lowk}$ than for the BHF approximation (see
Fig.~\ref{fig:effMass}). Results on effective masses obtained from SCGF are
rather similar to the BHF results, therefore we do not discuss them here
explicitly.

We want to add that the results for the effective $E$-mass can be rather
different in finite nuclei than in nuclear matter. As it has already been
mentioned above, this $E$-mass is due to the admixture of vibrational modes,
which are quite different in finite nuclei as compared to nuclear matter. This
may explain the differences in the predictions for the effective isovector mass
resulting from Skyrme interactions, which are based on fits of properties for
finite nuclei, and those originating from realistic interactions.  
  
\begin{figure}
\begin{center}
\mbox{
\includegraphics[width =10cm]{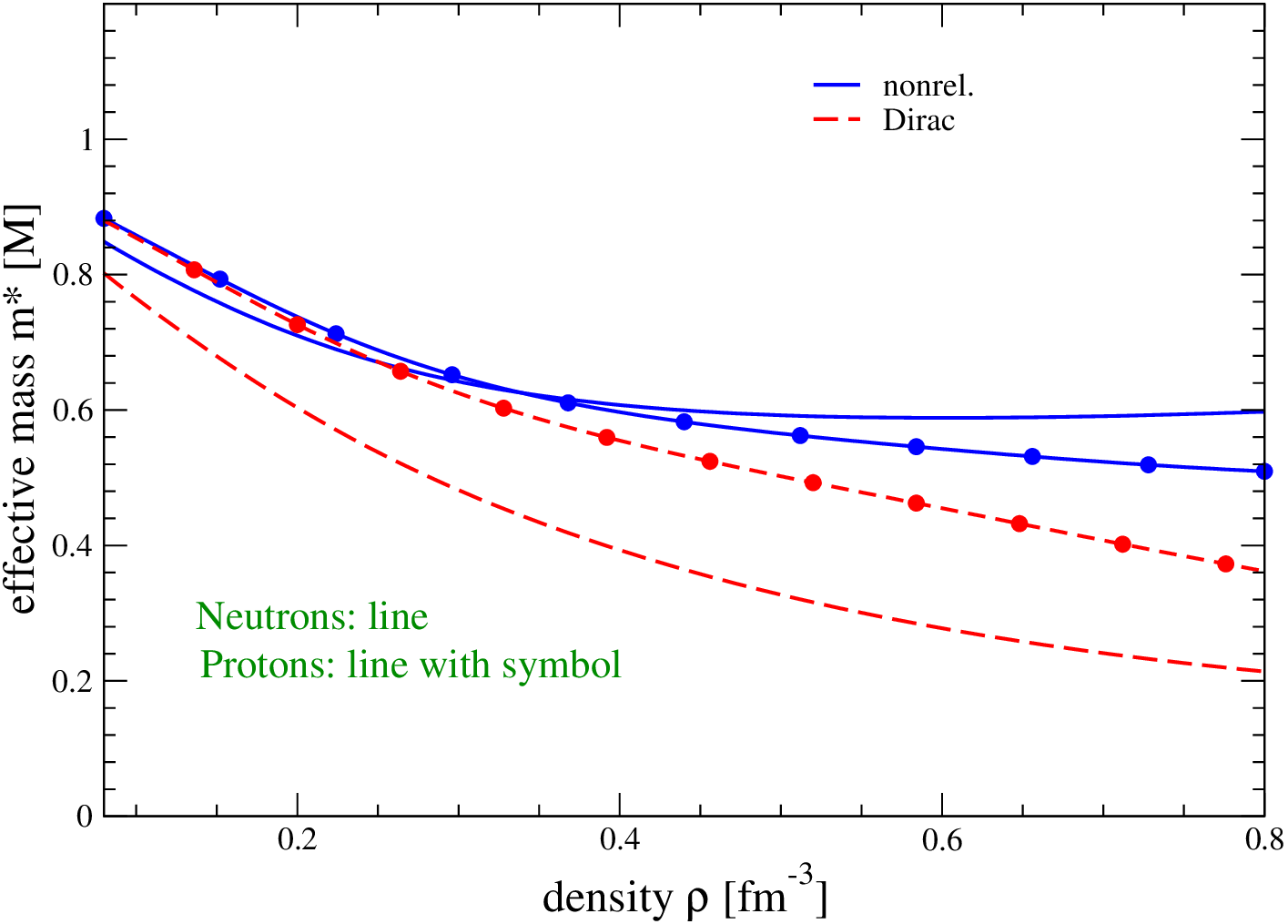}
}
\end{center}
\caption{\label{fig:MeffDDRMF} (Color online) Effective masses originating from
the DDRMF calculations of nuclear matter in $\beta$-equilibrium. The Dirac masses
for protons and neutrons are represented by the dashed lines, while solid lines
are used to identify the effective masses according to
eq.(\protect{\ref{eq:defmas}}). The results for protons are shown in terms of
lines with symbols.}
\end{figure}

Finally we are going to address the results for effective masses as they
originate from relativistic mean-field approximations. Here we have to
distinguish between the values for the Dirac mass and the effective mass
parameterizing the single-particle spectrum according to eq.(\ref{eq:defmas}).
The Dirac mass $m_D^\ast$ is defined in terms of the scalar part of the 
self-energy (see eq.(\ref{eq:defrels})). In the case of the relativistic
mean-field approximation the self-energies do not depend on energy or momentum
and the scalar self-energies for protons and neutrons are solely due to the 
direct contributions of the scalar mesons $\sigma$ and $\delta$. 
The Dirac masses obtained from DDRMF in nuclear matter at $\beta$-equilibrium
are displayed in Fig.~\ref{fig:MeffDDRMF}. We see that the Dirac masses decrease
with density but show larger values for the protons than for the neutrons.

In order to compare these results for the Dirac mass with the corresponding
non-relativistic effective mass, we have to compare the expressions for the
single-particle energies eq.(\ref{eq:dirace}) and eq.(\ref{eq:defmas}) and
adjust the parameters in such a way that the expressions yield identical results
and slopes as a function of $k^2$ at the Fermi momentum. This leads to
\begin{equation}
  m^\ast_i = \sqrt{k_{Fi}^2+(m_{Di}^\ast)^2}\,,\label{eq:diffmas}
\end{equation}
where the label $i$ refers to the case of proton and neutron. Results for these
non-relativistic masses are displayed in Fig.~\ref{fig:MeffDDRMF} by solid
lines. We find that the enhancement of the Dirac mass by the corresponding Fermi
momenta in eq.(\ref{eq:diffmas}) is significant for the neutrons in particular.
This leads to the effect that at high densities the non-relativistic effective 
mass for neutrons gets larger than the corresponding mass for protons, a
behavior which is opposite to the one observed for the Dirac masses.

It is worth noting that this feature, the difference of the Dirac masses
$m_{Dn}^\ast - m_{Dp}^\ast$ is negative while the difference of the
corresponding non-relativistic masses is positive has been observed before
within the framework of Dirac Brueckner Hartree Fock (DBHF) even at small
densities\cite{Schill01,vD07,Sama05,vvDD05}. The parameterization of the DDRMF approach
has been made to reproduce the bulk properties of the DBHF of \cite{vD07}.
However, this adjustment cannot account for details like the non-locality of the
self-energies in DBHF. Therefore it does not reproduce such details as the
effective isovector mass with good accuracy.

\section{Conclusion}

Various approaches to the nuclear many-body problem have been
investigated to explore their predictions for nuclear matter at high density and
large proton - neutron asymmetries. Two of these approaches, the Skyrme
Hartree-Fock and the Density Dependent Relativistic Mean Field approach are
predominantly of phenomenological origin. Their parameters have been adjusted to
reproduce data of finite nuclei. However, the parameters have been selected in
such a way that also bulk properties of asymmetric nuclear matter derived from
microscopic calculations are reproduced. The other three approaches are based on
realistic NN interactions, which fit the NN scattering phase shifts. In these
approximation schemes (Brueckner Hartree Fock BHF, Self-consistent Greens
Function SCGF and Hartree Fock using a renormalized interaction $V_{lowk}$) a
isoscalar contact interaction has been added to reproduce the empirical
saturation point of symmetric nuclear matter. 

These various approximation schemes lead to rather similar predictions for the
energy per nucleon of symmetric and asymmetric nuclear matter at high densities.
In detail one finds that the relativistic DDRMF leads to a rather stiff Equation
of State (EoS) for symmetric matter while the BHF approach leads to a relatively
soft EoS, a feature which is compensated within the microscopic framework by the
repulsive features of the hole-hole ladders included in SCGF. The
phenomenological approximation schemes DDRMF (Skyrme Hartree Fock) over (under)
estimate the symmetry energy at high densities as compared to the microscopic
approaches. The lack of long range (low energy) correlation effects in
$V_{lowk}$ leads to a symmetry energy which is too small already at normal
density. These features are also reflected in the study of nuclear matter in the
$\beta$-equilibrium and lead to moderate differences in the predictions for proton
abundances and EoS.

More significant differences are observed when we inspect details like the
effective masses, in particular the isovector effective mass. In neutron-rich
matter the microscopic approaches predict a positive difference between
neutron and proton effective masses. This feature can be related to the
non-locality of the self-energy induced by one-pion exchange term and is
expressed in terms of an effective $k$-mass. This feature may partly be
compensated by the effects of vibrational excitation modes on the nucleon mean
fields. The effects of such low-energy excitations might lead to different
results in nuclear matter and finite nuclei. This could be study e.g. in
many-body calculations employing $V_{lowk}$, which account for the effects of
vibrational modes explicitly.

We also discuss the differences between effective Dirac masses and corresponding
non-relativistic masses in neutron-rich matter. While the difference between the
neutron and proton Dirac masses is negative, the differences of the corresponding
non-relativistic masses tend to get positive. Further studies of the
non-localities in space and time of various components in the Dirac self-energy
would be useful to explore the connections to the non-relativistic microscopic
approaches more in detail.

\section{Acknowledgements}
This work has been supported by the Deutsche Forschungsgemeinschaft DFG (Mu
705/5-1).

\end{document}